\documentclass[aps,prl,twocolumn,a4paper,superscriptaddress]{revtex4}

\usepackage[paperwidth=215mm,paperheight=300mm,centering,hmargin=1.6cm,vmargin=2cm]{geometry}
\usepackage{graphicx,amsmath,SIunits}
\usepackage{float}
\usepackage{latexsym,stmaryrd}
\usepackage[sort&compress]{natbib}
\usepackage{color}
\usepackage{amsmath}

\begin{document}

\title{A self-assembled two-dimensional thermo-functional material for radiative cooling.}

\author{J. Jaramillo-Fern\'{a}ndez}
\email{juliana.jaramillo@icn2.cat}
\affiliation{Catalan Institute of Nanoscience and Nanotechnology (ICN2), CSIC and BIST, Campus UAB, Bellaterra, 08193 Barcelona, Spain}
\author{G. L. Whitworth}
\affiliation{Catalan Institute of Nanoscience and Nanotechnology (ICN2), CSIC and BIST, Campus UAB, Bellaterra, 08193 Barcelona, Spain}
\author{J. A. Pariente}
\affiliation{Materials Science Factory, Instituto de Ciencia de Materiales de Madrid (ICMM), c/Sor Juana In\'{e}s de la Cruz 3, 28049 Madrid, Spain}
\author{A. Blanco}
\affiliation{Materials Science Factory, Instituto de Ciencia de Materiales de Madrid (ICMM), c/Sor Juana In\'{e}s de la Cruz 3, 28049 Madrid, Spain}
\author{P. D. Garc\'{i}a}
\affiliation{Catalan Institute of Nanoscience and Nanotechnology (ICN2), CSIC and BIST, Campus UAB, Bellaterra, 08193 Barcelona, Spain}
\author{C. L\'{o}pez}
\affiliation{Materials Science Factory, Instituto de Ciencia de Materiales de Madrid (ICMM), c/Sor Juana In\'{e}s de la Cruz 3, 28049 Madrid, Spain}
\author{C. M. Sotomayor-Torres}
\affiliation{Catalan Institute of Nanoscience and Nanotechnology (ICN2), CSIC and BIST, Campus UAB, Bellaterra, 08193 Barcelona, Spain}
\affiliation{ICREA - Instituci\'o Catalana de Recerca i Estudis Avan\c{c}ats, 08010 Barcelona, Spain}

\small

 \pacs{(42.25.Dd, 42.25.Fx, 46.65.+g, 42.70.Qs)}

\maketitle

\textbf{The regulation of temperature at the macro and microscale is a major energy-consuming process of humankind.\ Modern cooling systems account for $15\,\%$ of the global energy consumption and are responsible for $10\,\%$ of greenhouse gas emissions.\ Due to global warming, a ten-fold growth in the demand of cooling technologies is expected in the next 30 years \cite{Goldstein2017, Future}, thus linking global warming and cooling needs through a worrying negative feedback loop.\ Here, we propose an inexpensive solution to this global challenge based on a single-layer of silica microspheres self-assembled on a soda-lime glass substrate. This two-dimensional (2D) crystal acts as a visibly translucent thermal blackbody for above-ambient daytime radiative cooling and can be used to passively improve the thermal performance of devices that undergo critical heating during operation.\ The temperature of a crystalline silicon wafer was found to be $14\, \text{K}$ lower during daytime when covered with our thermal emitter, reaching an average temperature difference of $19\, \text{K}$ when the structure was backed with a silver layer.\ In comparison, the soda-lime glass used as a reference in our measurements lowered the temperature of the silicon wafer by just $5\, \text{K}$. The cooling power density of this rather simple radiative cooler under direct sunlight was found to be up to $ \text{350  W}/\text{m}^{2}$ when applied to hot surfaces with relative temperatures of 55 K above the ambient. This is crucial to radiatively cool down electronic devices, such as solar cells, where an increase in temperature has drastic effects on performance.\ Our 2D thermo-functional material includes a single layer of spheres emitting long-wave radiation through the infrared atmospheric window over a broad wavelength range, thus providing effective radiative cooling using the outer space as a heat sink at $3\,\text{K}$.}

\

All modern cooling technologies are overwhelmingly energy-consuming.\ Air conditioners and refrigerators together account for almost a quarter of the total electricity used in the United States~\cite{Department}.\ Thus, it is not surprising that the latest International Energy Agency report~\cite{Future} identifies cooling as one of the main drivers of global electricity demand over the coming decades, aggravated in particular by global warming.\ This results in a negative feedback mechanism as cooling technologies already account for $15\,\%$ of the global electricity consumption and $10\,\%$ of greenhouse gas emissions~\cite{Goldstein2017}, figures predicted to triple by 2050~\cite{Future}.\ In addition, the efficiency of non-fossil alternatives such as solar photovoltaics~\cite{Vaillon2018} or nuclear power often relies on their operation temperature.\ Efficient heat removal in such systems will translate into increased performance and global energy savings.\ Modern cooling approaches thus require a fundamental rethinking in terms of environmental and economical sustainability, to reduce their impact on the global energy use.
 \begin{figure*}[t!]
	\centering
	\includegraphics[width=16cm]{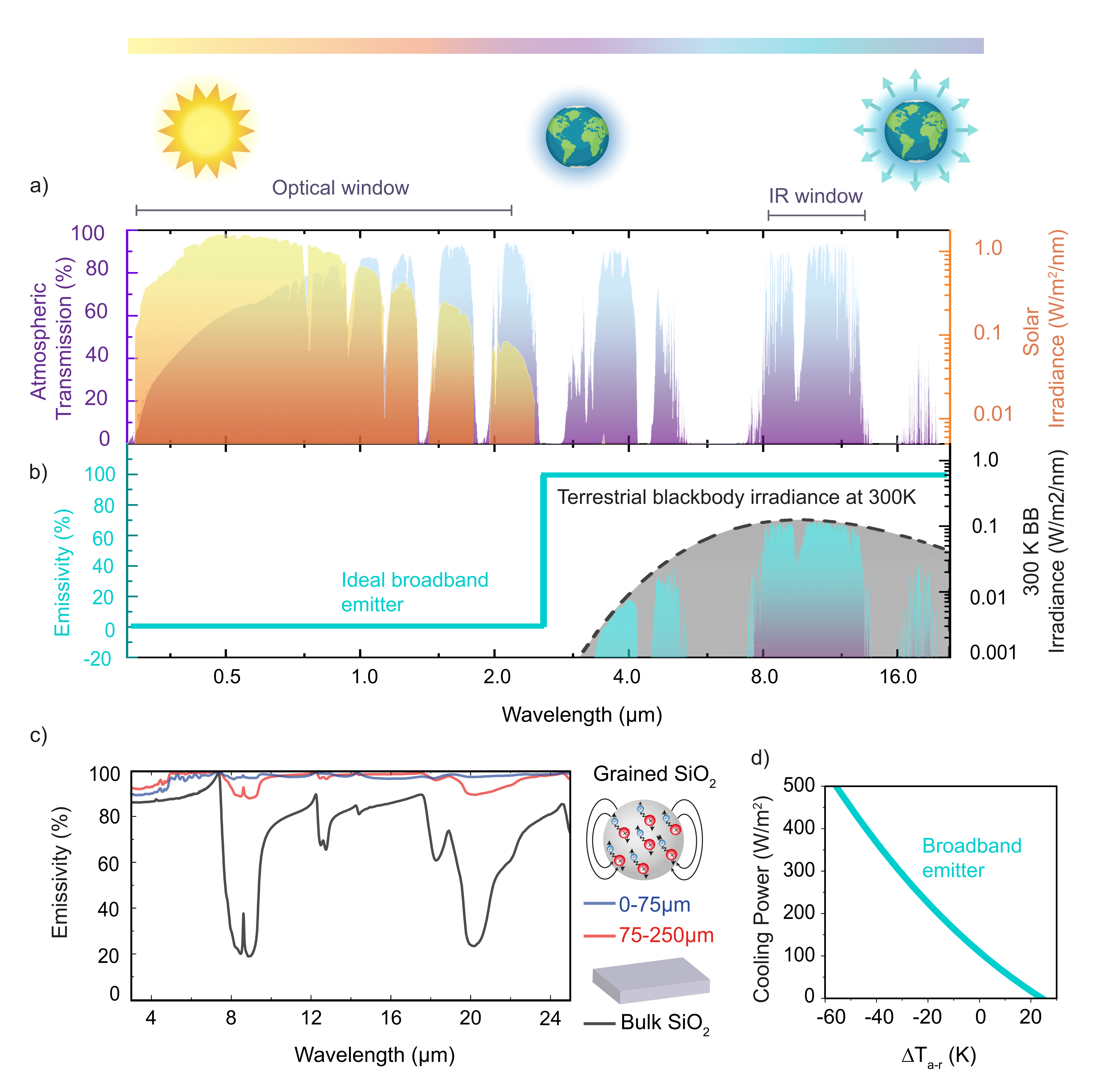}
	\caption{ \label{1} \textbf{Ideal thermal infrared blackbody for radiative cooling.} (a) The solar irradiance A.M 1.5 (reference spectrum at mid-latitudes with an air mass coefficient of 1.5) \cite{ASTM2012} and the atmosphere transmission calculated for Barcelona with MODTRAN \cite{Berk2014} (b) Spectral response of an ideal broadband radiator (continuous cyan line), reflecting or transmitting visible and near-IR light up to $2.5\,\mu \text{m}$ and re-emitting as a blackbody in the infrared spectral range. The terrestrial blackbody irradiance (at 300K) is also plotted for comparison.\ For the sake of clarity, the top color-bar relates the color-code used for the solar irradiance, the atmospheric transmittance and the terrestrial blackbody irradiance spectra to a schematic representation of the Earth's temperature-regulation mechanism: radiative cooling. The Sun's irradiance is filtered through the optical transmittance window of the atmosphere and is  then absorbed by the Earth's surface, rising its temperature.\ A portion of this energy runs the climate's heat engine while the rest is radiated back to outer space as infrared thermal energy through the IR atmospheric window. (c) Emissivity spectra of bulk (black line) and granular (red and blue lines) silica, extracted from reflectivity data of the ASTER library for particle sizes from 75-250$\,\mu \text{m}$ and 0-75$\,\mu \text{m}$, respectively.\cite{data} \ (d) Net radiative cooling power density calculated for a perfect broadband radiator considering relative temperatures between the ambient, $T_{a}$, and the radiator surface, $T_{r}$, from -60 to 30 K.\ }
\end{figure*}

Radiative cooling is a passive thermodynamic mechanism through which the temperature of the Earth is kept stable~\cite{Stocker2014}.\ Under ordinary conditions (neglecting \textit{anthropogenic forcings}) our planet is in radiative equilibrium, a state of nearly perfect balance between the incoming and outgoing energy. This energy exchange between the Earth, the Sun and the outer space takes place through the Earth's atmosphere, enabled by its radiative properties. \ To illustrate this, the atmospheric transmission and the solar irradiance (1.5 A.M spectral reference \cite{ASTM2012}) are plotted in Fig.~\ref{1}a.\ It shows how the Sun's irradiance is filtered through the optical transmittance window of the atmosphere, principally at visible and near-infrared wavelengths ($0.3\,\mu \text{m}$ to $2.5\,\mu \text{m}$).\ This radiation is then absorbed by the Earth's surface, rising its temperature.\ A portion of this energy runs the climate's heat engine while the rest is radiated back to outer space as infrared thermal energy in the range between $10$ to $13\,\mu \text{m}$ (IR window) in amounts proportional to $T^4$, thus being a process very sensitive to the temperature. Interestingly, the peak emission of a blackbody at average ambient temperature ($T_{a}$= 300 K e.g the Earth) perfectly coincides with this IR window.  This opto-thermal mechanism keeps the temperature of our planet stable,  making use of naturally occurring \textit{green} non-toxic materials, such as  sand, primarily made of silica ($\text{SiO}_2$).\ In fact, the sand  in deserts plays a major role in this process by reflecting sunlight and radiating thermal energy to the outer cold space.\ The energy radiated is directly proportional to its emissivity, which in turn depends strongly on the geometry, size and surface of sand grains.\ A comparison of the emissivity of granular silica for two ranges of particle sizes~\cite{data} and bulk silica is shown in Fig.~\ref{1}c.\ Bulk silica exhibits emissivity dips at around $8\,\mu \text{m}$ and $20\,\mu \text{m}$, arising from the interaction of optical phonons and electromagnetic radiation, so-called bulk phonon-polariton resonances, which result in high reflectivity (low emissivity) spectral regions, something rather detrimental for radiative cooling.\ In contrast, granular silica exhibits high broadband emissivity over the full infrared spectrum, which increases with decreasing particle size until it nearly reaches unity for diameters below $75\,\mu \text{m}$.\ Such emissivity response closely resembles that of an ideal infrared blackbody (cyan continuous line in Fig.\ref{1}b), revealing the enhanced radiative cooling properties of sand grains as compared to their bulk counterpart.

\ Earth's efficient temperature-regulation mechanism relies on this cooling principle and its understanding opens up an incredibly promising possibility of using microstructuring of polar materials to achieve energy-efficient heat removal. For instance, attempts to exploit this concept for terrestrial thermal management have been done in the past~\cite{Orel1993,Catalanotti1975,Gueymard2002,Granqvist1981}, yet it was not until very recently that passive cooling under direct sunlight was achieved for a silicon absorber using a visibly transparent broadband thermal blackbody~\cite{Zhu2013}.\ Daytime cooling is challenging because even a small amount of visible light absorption by the structure results in solar heat gain, which is detrimental for radiative cooling. The daytime cooling solution proposed by Zhu \textit{et al.} consisted of a periodic silica photonic structure fabricated by photo-lithography  Since then, other designs including photo-lithography-based photonic microstructured surfaces~\cite{Raman2014,Zhu2014,Zhou2016,Hossain2015}, polar mutilayered structures~\cite{Raman2014,Herve2018}, coatings~\cite{Gentle2015,Peng2018} and polar-polymer metamaterials on metal mirrors have been proposed ~\cite{Zhai2017,Mandal2018, Cai2018, Gentle2010a}, achieving  radiative cooling power densities higher than 90 W/m$^2$.\ Very recently, using a glass-polymer hybrid $80\,\mu \text{m}$-thick structure composed of randomly positioned silica spheres encapsulated by a transparent polymer,  a $93\,\text{W}/\text{m}^{2}$ cooling power density was achieved under direct sunlight~\cite{Zhai2017}.
In this work, we present a radiative cooler for above-ambient daytime passive cooling based on a self-assembled two-dimensional array of silica microspheres on a soda-lime substrate.\ With our thin thermo-functional material, we achieve a difference of 14 K for a silicon wafer when the thermal emitter is placed on its surface, which  to the best of our knowledge is the highest value achieved so far for above-ambient cooling under such a configuration. The achieved temperature decrease corresponds to an estimated cooling power density from  $ \text{125 W}/\text{m}^{2}$ to $ \text{107  W}/\text{m}^{2}$ at the above-ambient steady state operation temperature. We also measured the average radiative cooling power above-ambient temperature to
be between $\,\text{160 W}/\text{m}^{2}$ to $ \text{350  W}/\text{m}^{2}$ for values of relative temperatures $\Delta T_{a-r}$ ranging from -25 to -50 K, with $T_{r}$  being the operating temperature of the cooler. This is the first time that radiative cooling performance of hot surfaces has been studied, which is important when applying radiative cooling technologies to devices that suffer from critical heating efficiency roll-off. Furthermore, we show that only a single-layer of microspheres is necessary to achieve the highest cooling performance, which greatly relaxes the material constrains and reduces the costs required to mass-produce an optimal cooling structure.

\


\

The figure of merit of a generic radiative cooler at an operating temperature $T_{r}$ its given by the net cooling power, $P_\text{net}$, which  is the sum of incoming and outgoing energy per unit time in the system~\cite{Granqvist1980}. When the heat loss from the emitting surface is greater than the heat gain, there is a net cooling effect. This net cooling power for a given wavelength and angle of incidence, is defined as the difference between the spectral power density of IR radiation emitted from the surface through the atmosphere, and that of the absorbed incident light from the environment (both ambient blackbody emission and solar radiation).
This quantity is therefore directly linked to the spectral and angular emissivity of the radiative cooler, $\varepsilon_{r}(\lambda,\theta)$:
	%
	\begin{equation}
	\label{power}
	P_{net} (T_{r}, T_{a})= P_{r}(T_{r}) - P_{a}(T_{a}) - P_{sun} - P_{non-rad}(T_{r},T_{a})
	\end{equation}
	%
%

\
 In Eq.\ref{power}, the cooler radiative power, $P_{r}(T_{r})$, is thus given by: 
\begin{equation}
\label{Pr}
P_{r}(T_{r})=2\pi\int_{0}^{\pi/2}\int_{0}^{\infty}I_{BB}(T_{r},\lambda)\varepsilon_{r}(\lambda,\theta)\sin(\theta)\cos(\theta)d\lambda d\theta
\end{equation}
 with $I_{BB}(T_{r},\lambda)$ being the spectral blackbody  irradiance at the emitter operation temperature and $\theta$ the zenith angle. 
In Eq.\ref{power}, the absorbed power due to the atmospheric thermal radiation is calculated as:  
\begin{equation}\label{Pa}
\begin{aligned}
P_{a}(T_{a})=
&2\pi\int_{0}^{\pi/2}\cos(\theta)\sin(\theta)d\theta\times\\
&\int_{0}^{\infty}I_{BB}(T_{a},\lambda)\varepsilon_{a}(\lambda,\theta)\varepsilon_{r}(\lambda,\theta)d\lambda 
\end{aligned}
\end{equation}

and the corresponding term due to sunlight absorption by the radiative cooler is $P_{sun}=\int_{0}^{\infty}I_{AM\ 1.5}(\lambda)\varepsilon_{r}(\lambda,\theta_{sun})d\lambda$, with $I_{AM \ 1.5}  $ being the reference solar irradiance 1.5 AM.\cite{ASTM2012}. Note that the integral over the solid angle is not necessary here since we assume the radiative cooler to be facing the Sun. Finally, $P_\text{non-rad}=h(\Delta T_{r-a})$ is the parasitic non-radiative contribution due to the energy exchange with the surrounding media via convection and conduction, with $h$ being a thermal coefficient that captures the non-radiative losses.\
Using Eq.~\ref{power}, we calculate the effective radiative cooling power density of an ideal broadband infrared emitter as a function of the relative temperature between the ambient, $T_{a}$, and the radiator, $T_{r}$. Here, we use the term of ideal broadband infrared emitter to refer to a material that has unity emissivity in the infrared spectrum beyond 2.5 $\micro$m, in contrast to a selective cooler that emits strongly only in the infrared atmospheric window (8 to 13 $\mu$m).\ The spectral profile of an ideal broadband radiator and its corresponding net radiative cooling power are shown in Fig.~\ref{1}b   (continuous cyan line) and Fig.~\ref{1}d, respectively. Figure ~\ref{1}d shows that, in agreement with previous studies ~\cite{Sun2017,Zhu2015,Hossain2016,Kou2017},  broadband radiators are optimal for above-ambient cooling, where the desired temperature is comparable to or higher than the ambient temperature. In contrast,  selective coolers perform better for applications below-ambient temperature. Thus, broadband emitters have greater cooling performance during daytime than their selective counterparts if the surface that needs to be cooled down is at higher temperature than the ambient $(T_{r}>T_{a})$. For instance, broadband emitters can evacuate heat to the outer cold space with a power density up to $400\,\ W/m^2$ at $\Delta T_{a-r}=-40\,\ K$ (Fig.\ref{1}d). This represents, for example, about 62 $\%$ of the power dissipated as heat by a solar cell. Passively removing this amount of heat would translate into a relative efficiency increase  up to 10 $\%$, since solar cell efficiency is reduced by 0.5 $\%$ for every degree of temperature increase~\cite{Gueymard2002}.

\
\begin{figure*}[t!]
	\centering
	\includegraphics[width=16cm]{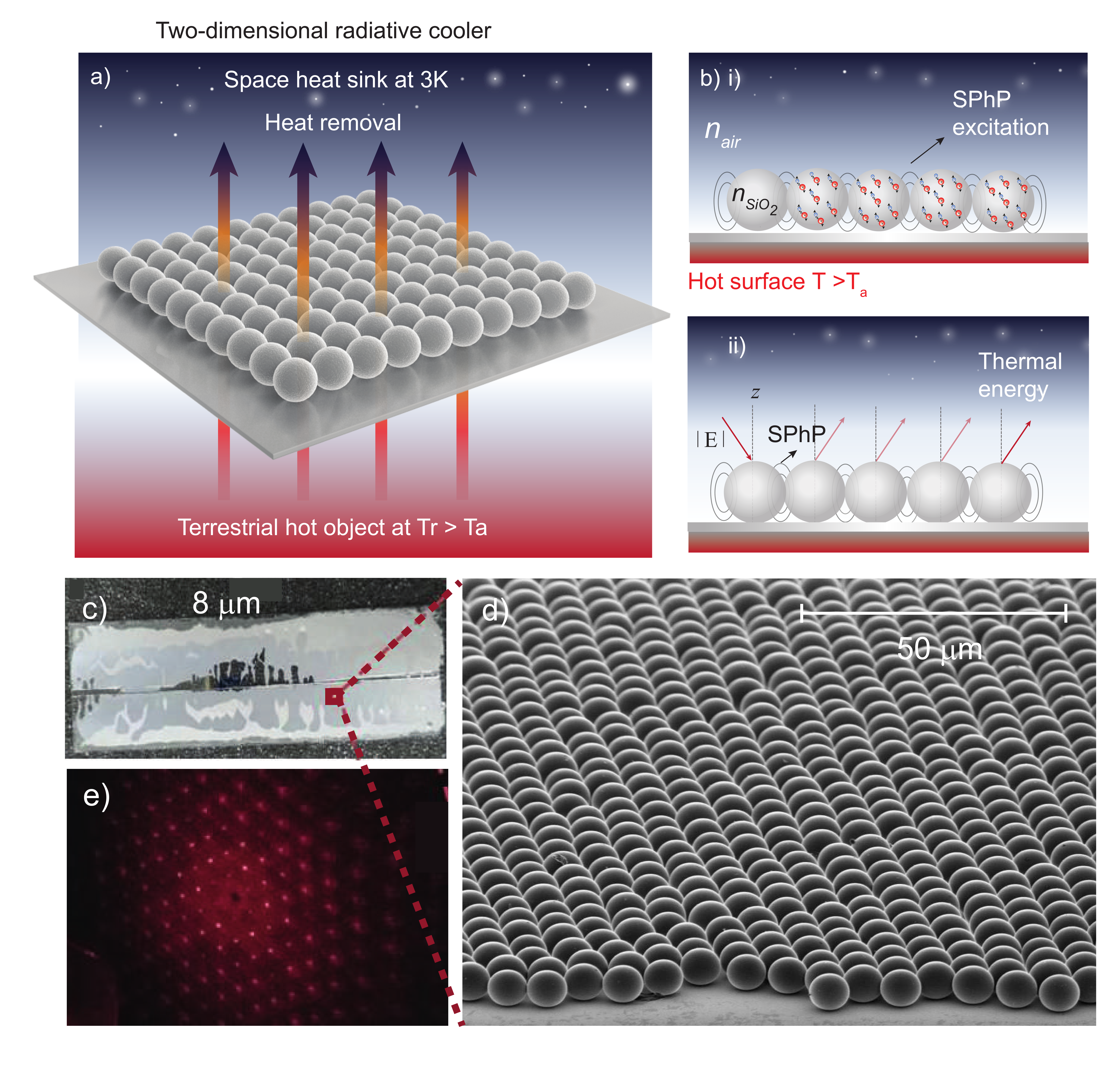}
	\caption{ \label{2} \textbf{Self-assembled two-dimensional crystal as a near-blackbody in the thermal infrared.} (a) Schematic representation of (a) the self-assembled 2D radiative cooler and  (b) its working principle : (i) Localized surface phonon polaritons are thermally excited in the colloidal crystal, which is integrated onto the hot surface of a sensitive device that undergoes severe heating ($T_{r}>>T_{a}$). The  SiO$_{2}$ microsphere array provides an interlayer with packing fraction of 30\% and the shape of the microspheres results in an effective impedance matching between the polar material and the dielectric medium, contributing to the high infrared emissivity of the radiative cooler. ii) Heat is radiated as thermal energy due to the enhanced emissivity resulting from the gradual change in shape and the diffraction of surface phonon polaritons via the 2D self-assembled grating. (c) Optical image and (d) scanning electron micrograph of a microscope slide coated with a colloidal crystal made of $8\,\mu \text{m}$-diameter spheres.\ (e) Laser diffraction pattern for red incident light on the single-layer colloidal crystal.}
\end{figure*}
\

Our approach to maximize the radiative cooling figure of merit given by Eq.\ref{power} rests on the use of a self-assembled 2D crystal composed of $8\,\mu \text{m}$-diameter silica spheres.
The simple radiative cooler that we report here acts as a nearly-ideal infrared broadband emitter that preserves a high transmittance to visible light.  Our thin thermo-functional material is grown directly on a $1\,\text{mm}$-thick soda-lime glass slide substrate (Fig.\ref{2}a).\ The high homogenity in size and shape of the microspheres (8.0 $\pm \ 0.2 \ \micro$m) allows them to self-organize into dense close packed periodic structures.\ A schematic illustration of the structure is depicted in Fig.~\ref{2}a, while an optical image and a scanning electron micrograph are shown in Fig.~\ref{2}c and Fig.~\ref{2}d, respectively.\ In our samples, this single-layer colloidal crystal typically covers $94\,\%$ of the substrate ($25\,\text{mm} \times 75\,\text{mm}$), as shown in Fig.~\ref{2}c.\ Figure \ref{2}e reveals the long-range ordering of the structure with large crystal domains of hundreds of micrometers in average (see Supplementary Material and Fig. S\ref{1}).\ The slight differences in crystal grains orientation are mostly due to vacancies or dislocations induced during the fabrication process and are believed to have negligible effect on the cooling functionality or even enhance it thanks to directional scattering \cite{Moyroud2017, Atiganyanun2018}.\ We test the high quality of our structures over an area of few $\text{mm}^2$ by optical diffraction of a normal-incidence red laser (Fig.~\ref{2}e) where the six defined spots correspond to an hexagonal lattice.\ The 2D colloidal system can be mass-produced to allow for a $\text{cm}^2$-scale technologically ready approach for radiative cooling~\cite{Zhao2016}.\ This structure behaves as a nearly-perfect broadband infrared thermal emitter, exhibiting an infrared emissivity greater that $98\,\%$ at ambient temperature, while being translucent to the visible light due to the scattering properties of its structural periodicity and grain defects (see suplementary material for further details).\ Ultra high infrared emissivity is achieved by combining three effects i) the spherical shape of the SiO$_2$ beads, ii) the thermal excitation of localized hybrid surface modes, i.e., surface phonon polaritons (SPhP) and iii) the diffraction of such excitations to the far field via the periodic lattice. The three mechanism are illustrated in Fig.~\ref{2}b and described here in detail.

\
i) The differential polar-dielectric boundaries from the microspheres result in a progressive variation of the material effective refractive index, $n_{eff}=c/(d \mathbf{\omega}/d \mathbf{\kappa})$ \cite{Lopez2006}, which is a function of wavelength and microsphere size, with $\mathbf{\omega}$ and $\mathbf{\kappa}$ being the energy and the wavevector of the photons.  This yields an impedance matching between the polar and the dielectric media (silica/air) over a large spectral range~\cite{Zhu2015, Zhu2014,Zhai2017}.  Moreover, the $8\,\mu \text{m}$-diameter spheres act individually as resonators, where high-order Fr\"{o}hlich resonances  are excited \cite{Zhai2017,Frohlich1949}.  

\ 
ii) The undesired temperature rise generated by dissipation in the surface that requires cooling i.e a solar absorber, becomes a heat source to activate localized SPhP resonances in the polar spheres~\cite{Joulain2005}. SPhP are collective mechanical oscillations of the dipoles in the polar sphere that interact resonantly with photons of the same frequency. At the surface, this results in an evanescent field with high intensity and strong near field confinement. However, the effect of this surface modes is not transmitted in the far field.

iii) By ruling a grating with period $d$ on the interface between the polar and the dielectric media, SPhP can be outcoupled into the far-field \cite{Greffet2002}, resulting in enhanced emissivity. Such SPhP-mediated emissivity depends on the polarization and the azimuth angle, $\theta$, according to $2 \pi \sin \theta / \lambda =\vec{q}_{x}+2\pi n / d$, where $n$ is an integer and $\vec{q}_{x}$ is the wavevector of the surface excitation. Our single-layer colloidal crystal acts as a grating in two-dimensions, where the crystal periodicity diffracts the SPhPs into the far field, thus further enhancing the emissivity for $p-$ and $s-$ polarizations, even at high incident angles (see Supplementary Material for further details). In this way, the generated heat is evacuated as thermal radiation optimized to pass through the transparent infrared atmospheric window via surface engineering.

\begin{figure*}[t!]
	\centering
	\includegraphics[width=16cm]{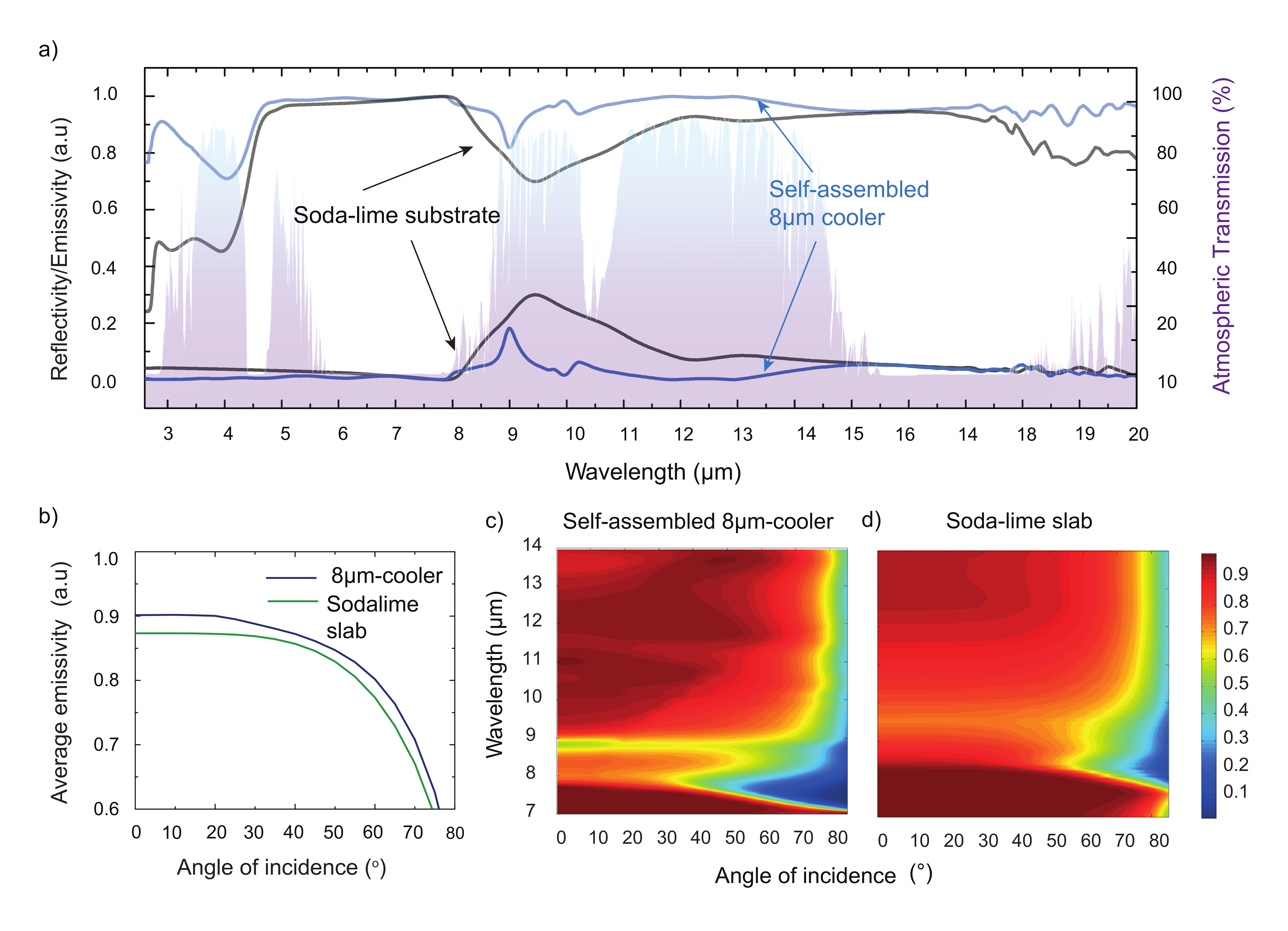}
	\caption{ \label{3} \textbf{Optical properties of a two-dimensional-crystal for radiative cooling.} (a) Reflectivity and emissivity spectra of soda-lime glass and single-layer radiative cooler made of $8\,\mu \text{m}$-diameter spheres.\ The MODTRAN~\cite{Berk2014} atmospheric transmittance model at Barcelona is plotted for comparison.\ (b) Thermal emissivity calculated by RCWA for the 2D-crystal cooler and a soda-lime reference substrate for angles of incidence from 0 to 80 degrees and averaged from $8\,\mu \text{m}$ to $13\,\mu \text{m}$.\  Wavelength- dependent unpolarized emissivity maps obtained by RCWA for incident angles varying from 0 to 85 degrees, corresponding to (c) the 2D-crystal thermal emitter and (d) the soda-lime reference slab, respectively.}
\end{figure*}

\

\begin{figure*}[t!]
 \centering
  \includegraphics[width=18.5cm]{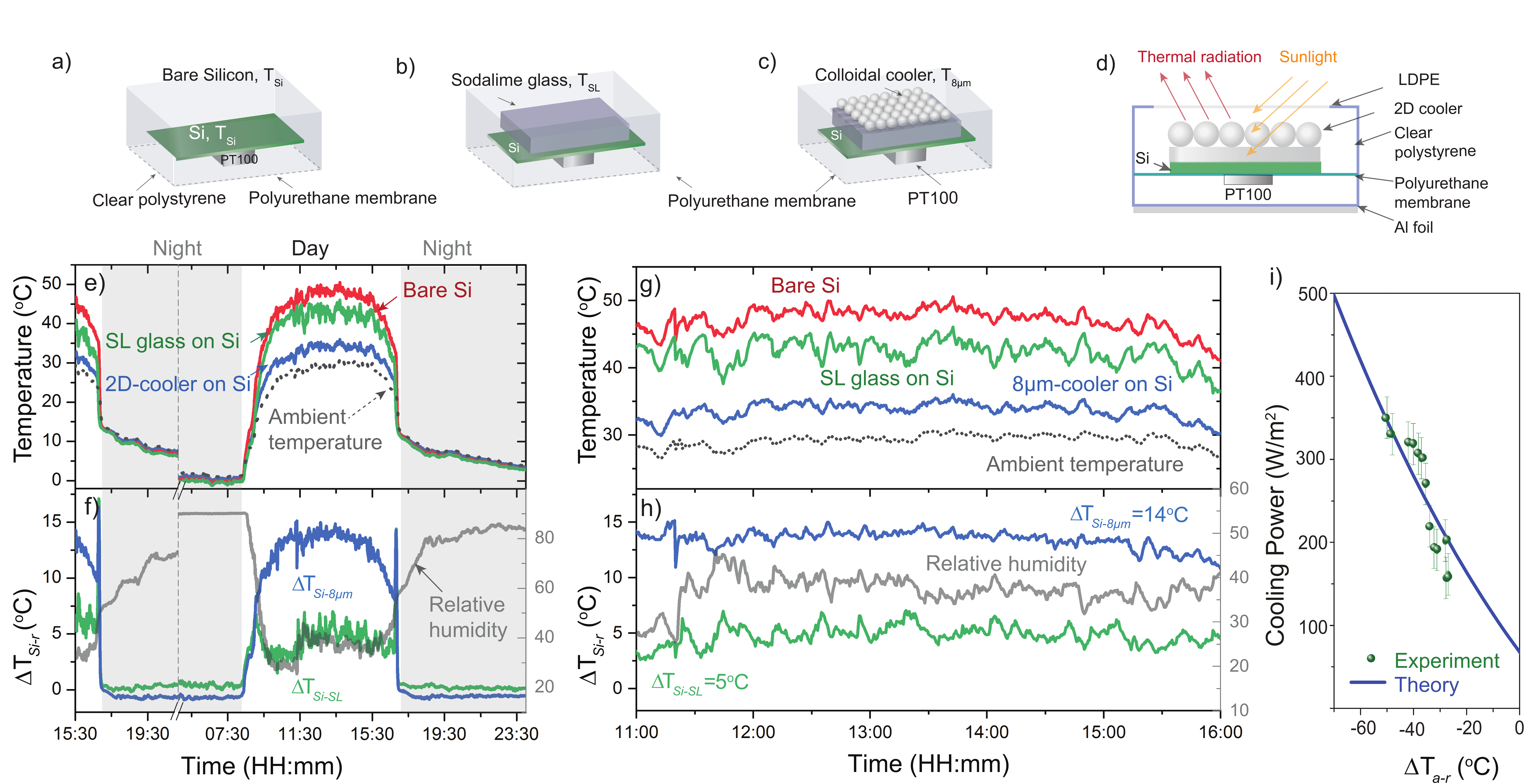}
    \caption{ \label{4} \textbf{Experimentally testing the radiative cooling.} Schematics of the test setups for (a) bare silicon substrate, (b) soda-lime (SL) glass on a silicon substrate (c) a 2D array of $8\,\mu \text{m}$-diameter spheres on soda-lime glass place on a silicon substrate.\ (d) The cross section of the 2D-crystal cooler and the test set-up for (c).\ Here, the thermal emitter is placed on a silicon substrate with a thermometer attached (see methods).\ The sample lies on a polyurethane membrane with low thermal conductivity, which is suspended inside a polystyrene box to minimize conduction losses and lateral convection.\ The polystyrene boxes are covered with a low-density polyethylene film that is transparent in all investigated wavelengths, in order to reduce convection though the top.\ The polystyrene boxes are supported by a wooden structure covered with aluminum foil to reflect the visible light and minimize parasitic heating.\ (e) Measured cycle of temperature during $32\,\text{hours}$ in a clear-sky day for the bare silicon substrate (red), the soda-lime cooled reference (green), and the  2D-crystal cooled Si (blue).\ The evolution of the ambient temperature is plotted as a black dotted line for reference \ (f) Temperature difference between the bare silicon and the colloidal-crystal cooler, as well as the soda-lime reference. The measured relative humidity is plotted on the right axis. Although atmospheric water vapor reduces the atmospheric transmittance and therefore reduces the radiative cooling efficiency, here the oscillations in $\Delta T_{Si-r}$ are mostly due to temperature variations and not to the actual moisture content of the atmosphere. Relative humidity changes slowly with time, and thus we assume that it does not introduce changes in the thermal conductivity and dissipation properties of the 2D material. Zoom-in of (g) the temperature measurement and (h) the temperature difference, $\Delta T_{Si-r}$ for the tested samples as they reach the steady state operation temperature.\ The temperature of bare silicon is passively lowered by an average of $14\, \degree \text{C}$ during sunlight hours with the self-assembled 2D-crystal cooler, compared to $ 5\degree \text{C}$ with the reference sample. (i) Radiative cooling power measured in response to a step-wise change in temperature of 5 $\degree$C set at the underlying surface where the cooler is placed using an on/off temperature controller. The values are obtained by comparing the current and voltage supplied on an identical set-up without the 2D-cooler. The two-dimensional crystal is able to evacuate heat with a cooling power density of 350 $W/m^2$, when $\Delta T_{a-r}=-50\degree$C.}
\end{figure*}

The high optical transmission and ultrahigh emissivity at thermal wavelengths of the singe layer array of microspheres provide the underlying thermo-optical mechanism responsible for passive cooling and determine its cooling power.\ To characterize the emissivity of our 2D cooler, we  measure the reflectivity, $R$, of the samples in the infrared spectrum (see Methods for details).\ At infrared wavelengths above $5 \ \micro\meter$~ \cite{Palik1998}, we can assume a negligible transmissivity $(T)$ for $1\,\text{mm}$ soda-lime substrate, so the fraction of radiation absorbed by the structure, the absorptivity A, can be obtained directly as $A=1-R$.\ This is an indirect approach to obtain the radiative emissivity as it is equal to the absorption in the corresponding spectral region under thermodynamic equilibrium, according to the Kirchhoff’s law of thermal radiation $(A=\varepsilon)$ \cite{Kirchhoff1860}.\ The measured reflectivity (and deduced emissivity) of a self-assembled 2D-crystal composed of $8\,\mu \text{m}$-diameter spheres is plotted Fig.~\ref{3}a.\ For reference, the corresponding curve of a soda-lime-glass slab is also shown.\ Soda-lime glass is  inexpensive, chemically stable and easy to melt and mold to make glassware and windows.\ In its bulk form, it acts as a fairly acceptable broadband radiative cooler material due to its weak  bulk phonon polarization resonance and high mid-to-long-infrared absorption.\ Thus, it is a perfect reference to compare with our 2D radiative cooler with an added perspective of its technological impact and future potential.\ At thermal wavelengths, a $1\,\text{mm}$-thick soda-lime substrate has a reasonable emissivity, $\varepsilon_\text{SL} = 0.85$, averaged over the IR atmospheric window, while our 2D-cooler exhibits an averaged thermal infrared emissivity $\varepsilon_\text{$8\mu$m cooler} > 0.98$.\ This is a remarkable value when compared to a perfect thermal blackbody, which has an emissivity of unity and a reflectivity of zero (from $2.5\,\mu \text{m}$ to $26\,\mu \text{m}$) and represents the ideal case to maximize the radiative cooling power at $\Delta T_{a-r}\leq 0$.\ Furthermore, the emissivity of our cooler improves over 0.93, a value obtained in Ref.~\cite{Zhai2017} using an $80\,\mu \text{m}$-thick randomized glass-polymer hybrid metamaterial i.e a 10 times thicker structure.\

The emissivity of our cooler is measured by FT-IR microscopy, using an objective with numerical aperture of 0.4, which can collect light at a solid angle up to 24$\degree$. To unreveal the directional diffraction properties of our self-assembled 2D-cooler, we model its optical response  over the full solid angle with a rigorous-coupled-wave analysis (RCWA) numerical code for this configuration (details in the Methods and Supplementary Material).\ The calculated emissivity averaged throughout the transparent infrared atmospheric window and for angles of incidence from 0 to 85 degrees, both for the single-layer crystal cooler and for the reference soda-lime glass substrate are shown in Fig.~\ref{3}b.\ The spectral emissivity of our 2D-cooler is $22\,\%$ higher than that of a soda-lime glass slab.\ The angular variation of emissivity from 0 to 85 degrees at the same spectral window shows a large enhancement of the cooler compared to the reference, even at high incident angles (Fig.~\ref{3}c and d, respectively).\ These results evidence the enhancement of the  emissivity mediated by SPhP diffraction by the periodic lattice, particularly in the infrared transparent atmospheric window where SiO$_2$ exhibits its phonon polariton resonances. Furthermore, the RCWA results show that the physical origin of the emissivity increase of our cooler is also due to the spherical geometry of the SiO$_2$ beads (see details in the Supplementary Material). \ In agreement with other studies, our model reveals no further improvement of the emissivity with increasing the number of colloidal layers ~\cite{Yuksel2017,Tervo2017} which greatly relaxes the material constrains required to achieve an optimum cooling mechanism.


To test the performance of our cooler, we designed an experiment in which the temperature of a silicon slab, $T_\text{Si}$, is compared to the temperature of the reference soda-lime glass slab, $T_\text{SL}$, or to the 2D-crystal cooler, $T_{8\,\mu \text{m}}$, when both are placed on the silicon wafer and exposed to direct sunlight.\ Silicon was used since it is the semiconductor that enabled modern electronic technologies, such as solar cells, and these would greatly improve their efficiency by operating at lower temperatures. Thus, standard 550 $\micro$m-thick boron-doped silicon wafers were  used as an example case, acting as solar absorbers.  Schematic illustrations of the studied structures in the experiment are shown in Fig.~\ref{4}a, b and c, respectively.\ In these different configurations, we measure the temperature of the reference, the thermal emitter and the silicon slab during clear sky exposure using thermometers attached to each system.\ Convection and conduction thermal losses as well as parasitic heating are minimized as detailed in the methods section and the Supplementary Material.\ Fig.~\ref{4}e shows a $32\,\text{h}$ temperature measurement for the three structures, under a clear sky and average relative humidity of $37\,\%$ and $74\,\%$ during the day and night, respectively.\ The silicon slab heats up during daylight hours due to sunlight absorption, reaching a maximum temperature of $51\,\degree \text{C}$ at 1:45 pm.\ Both the reference and the thermal radiative cooler keep the silicon slab colder, particularly during daytime, demonstrating above-ambient radiative cooling.\ Since the reference and the self-assembled 2D-cooler are highly transparent and translucent in the visible wavelength range, respectively, the upcoming sunlight is transmitted to the silicon slab, heating it up above the ambient $(\Delta T_{a-r}\leq 0)$.\   In this range, this broadband radiative cooler provides its highest achievable cooling power density, outperforming its selective counterparts~\cite{Hossain2016}.\ Around noon time, from $11\,\text{am}$ to $3\,\text{pm}$, average values of $\Delta T_{a-8 \mu \text{m}}$ from $-6\,\degree \text{C}$ to $-2\,\degree \text{C}$  were recorded, for which the corresponding net radiative cooling power ranges from $125\,\text{W}/\text{m}^{2}$ to $107\,\text{W}/\text{m}^{2}$.\ For the same values of $\Delta T_{a-r}$, the reference exhibits a considerably lower radiative cooling power ranging from $93\,\text{W}/\text{m}^{2}$ to $74\,\text{W}/\text{m}^{2}$ (Fig.S7 in the Supplementary Material).\ In addition, the average steady-state temperature difference measured during daylight hours is $14\,\degree \text{C}$ for our 2D-cooler and only  $5\,\degree \text{C}$ for the reference (Fig.~\ref{4}g and h, respectively).\\
The cooling power of both the reference and the passive cooler decreases as $\Delta T_{a-r}$ approaches zero, until they reach a steady state value of about $1\,\degree \text{C}$ below ambient temperature.\ The bare boron-doped silicon substrate also experiences this slight radiative cooling effect, that has been attributed in previous works to free-carriers infrared emission induced by the doping~\cite{Kou2017}.\ We achieve below-ambient radiative cooling and improve these numbers by placing our 2D-radiative cooler on a silver-coated silicon slab exposed to direct sunlight, which reflects most of the incident solar radiation, thus lowering the temperature of the silicon slab by $19\,\degree \text{C}$ with an operation temperature $T_{r}=2.5\,\degree \text{C}$ below the ambient one (Fig.S8 in the Supplementary Material).\
To corroborate our results, we measured the above-ambient radiative cooling power of the 2D-cooler using an alternative experiment, in which we obtain the power injected to a heater in response to a step-wise increase in temperature set at the underlying surface where the cooler is placed. The radiative cooling power is obtained by comparing the current and voltage supplied on an identical set-up without the 2D-cooler, measured simultaneously during 24 hours (See Methods and Supplementary Material). 

 The direct measurement of the radiative cooling power at $T_{a} < T_{r}$ is plotted in Fig.\ref{4}i, as a function of the relative temperature between the heater ($T_{r}$) and the ambient ($T_{a}$) and compared with the theoretical estimation (Eq.\ref{power}).\ The model includes the 2D-cooler typical absorption of 3\% in the solar spectrum and accounts for the convection and conduction losses. Equation \ref{power} yields remarkably satisfactory predictions of the cooler performance measured data, with an error below 10\%. 
 Figure \ref{4}i shows that our 2D-thermo-functional material can evacuate heat at a cooling power density between 160 and 350 $W/m^2$ for relative temperatures $\Delta T_{a-r}$ from $-25$ to $-50\degree$C. This represents from 40 up to 55\% of the heat dissipated by a solar cell under solar irradiance of 800 $W/m^2$ and indicates substantial gains in solar cell efficiency.  Considering the global solar energy production in 2017 \cite{IEA2018}, and assuming that the solar panel efficiency is 20\%, such an efficiency increase due to the radiative cooling would be enough to power a city of $2’374.000$, comparable to Paris, during an entire year. Moreover, the model shows that at ambient temperature $T_{a}=T_{r}$ , our cooler can evacuate heat at radiative cooling power densities of 70 $ W/m^2$, comparable to the values reported in references ~\cite{Hossain2016,Peng2018,Zhai2017, Herve2018}.
 
  Based on these values, our self-assembled 2D-cooler exhibit excellent thermo-optical and radiative cooling properties.\ In addition, given the $3\,\%$ absorption in the visible spectral range of the glass substrate used to grow our sample, we expect a further enhancement using  less-absorptive substrates like quartz.


We present here a novel thermo-functional material for passive radiative cooling inspired by the Earth's efficient temperature-regulation mechanism.\ It consists of a single-layer of a colloidal-crystal composed by SiO$_2$ spheres deposited on a soda-lime glass substrate.\ Our structure behaves as a nearly perfect thermal blackbody  with an emissivity higher than $98\,\%$ for wavelengths beyond $2.5\,\mu\text{m}$. We analyze its radiative cooling performance by measuring the temperature cycle of the proposed structure and estimating a cooling power density from  $ \text{125 W}/\text{m}^{2}$ to $ \text{107  W}/\text{m}^{2}$ at the steady state operation temperature. In addition, we prove its performance as an above-ambient daytime-cooler by measuring an average radiative cooling power from  $ \text{160 W}/\text{m}^{2}$ to $ \text{350 W}/\text{m}^{2}$ during night and day,  for relative temperatures, $\Delta T_{a-r}$, from -25 to -50 $\degree$. During a clear-sky day, this 2D-radiative cooler kept a silicon slab  $14\,\degree \text{C}$ cooler on average and $19\,\degree \text{C}$ when backed with a silver layer.\ The cooling performance of the proposed structure could be further enhanced by using a quartz substrate.\ The proposed cooling system is simple, extremely cost-efficient, light, very thin and avoids the use of plastics, which are aspects of high concern in current public opinion. In summary, the novel 2D material is a sustainable alternative for passive cooling at the macro and microscale.

\section{Experimental section}

\subsection{Sample design and fabrication}

A novel method [in preparation]  of sedimentation based on vertical deposition was used to grow the self-assembled 2D colloidal cooler. Briefly, two substrates were placed on a Petri dish, one of them lying at the bottom and the other one, forming $45\degree$  with respect to the former. This configuration of the substrates favors the assembly of the spheres through the long meniscus during the evaporation. Prior to deposition, hydrophilization was achieved by immersing the substrates of sodalime glass during 20 minutes in standard chromic acid solution which favors the self-assembly of the spheres. High quality photonic crystals with large area and controllable thickness can be built by this procedure (Figure 2c and S1 in the Supplementary Material). The concentration of the colloidal suspension is varied depending on the layers' thickness required and the size of the spheres. To achieve the formation of one monolayer of 8$\micro\meter$-silica spheres, 30mL of aqueous colloidal suspension (silica colloid concentration around 0.2 wt\%) was prepared and placed in a climatic chamber. The evaporation rate was controlled by keeping temperature and relative humidity to 45 $\degree$C and 20 \%, respectively, during 24 hours. Diffraction pattern, obtained with HeNe laser (Figure 2d) and SEM inspection (Figure 2e) reveal an extraordinary quality in the photonic crystal obtained by this method.

\subsection{Optical characterisation}

The reflectivity R of the sample in the infrared spectrum ($2.5 - 26 \ \mu$m) was measured by FTIR microscopy, using a Bruker Hyperion 2000 with an objective of 15 times magnifying power and a numerical aperture of $N.A=0.4$.\ The absorptivity was then obtained as $A=1-R$, due to the negligible transmission of SiO$_2$ at infrared wavelehgths ($T=0$) \cite{Palik1998}.\  The optical properties of the samples over the visible wavelength range, from 300 to 800 nm, were obtained using an ultraviolet/visible Cary 4000 spectrometer. We used integrating spheres to account for the scattered light from the full solid angle. Due to the transparency of SL glass and f-$\text{SiO}_2$, the absorption for these wavelengths is minimal i.e about 3.3\% averaged from 380 to 800 nm.

\subsection{Emissivity modelling}

For the electromagnetic modelling, an in-house built rigorous coupled-wave analysis code (RCWA), written in MATLAB, was employed to calculate the total transmission ($T$) and reflection ($R$) of the single-layer colloidal-crystal as a function of wavelength and angle of incidence.\ Absorption ($A$) is then calculated by the remainder of these two values from unity ($(A=1-T-R)$).\ To model an single-layer colloidal structure the RCWA assumed a unit-cell periodic in the $x$ and $y$ directions and discretised into analytical layers in the z-direction (as depicted in Fig.~S3).\ For the simulations shown in Fig.~3b-c each sphere of the colloidal self-assembled layer was discretised into 25 equally thick slices, with each slice being converted into Fourier space using 11 times 19 spatial harmonics.\ These simulations produced smooth continuous data sets as function of angle and wavelength and so the number spatial harmonics was judged to be sufficient.\ The SiO$_2$ colloidal-crystals were simulated atop a 1 mm-thick substrate of soda-lime glass to best emulate the experimental conditions \cite{Palik1998}.\ The atmospheric transmittance was obtained by using an open-source avaliable code (MODTRAN\textregistered \cite{Berk2014}) to calculate the spectral solar irradiance.

\subsection{Experimental setup for continuous temperature measurements}

In the continuous temperature experiment, the temperature of a silicon slab, $T_{Si}$, is compared to the temperature of the cooler, $T_{8\,\mu \text{m}}$, and to the reference soda-lime glass slab, $T_{SL}$, both placed atop the silicon slab and exposed directly to sunlight.\ The reference and the thermal radiator are index matched to the silicon slab using a commercial refractive index matching gel to avoid unwanted reflections due to the thin air layer that might otherwise form between the two surfaces.

The experiment simultaneously measures the temperature of the thermal emitters and the bare silicon slab during sky exposure using 4-point resistive thermometers (PT100) attached to the substrate surface with silver paste to ensure good thermal contact.\ To minimize the convection and conduction losses, the tested structures were placed on a chemically inert polyurethane membrane with a very low thermal conductivity.\ The polystyrene boxes, where the radiative cooler structures are placed, also serve to minimize conduction losses.\ Finally, the polystyrene boxes were covered with a low-density polyethylene thin film (0.01 mm) that has negligible absorption in all investigated visible thermal infrared wavelengths.\ The boxes are supported by a wooden structure that is covered with aluminum foil to reflect the visible light and avoid parasitic heating (see Fig.~S4b).\ Sunlight is normally incident on the radiative coolers during the duration of the measurement.\ This is achieved by attaching the wooden frame that supports the tested structures to a solar tracker based on a motorized rotating platform.\ We seek for normal incidence in our experiments, to estimate the performance of the radiative coolers under the toughest environmental conditions, for which sunlight absorption exhibits highest values (see Fig.~S4a).

\subsection{Radiative cooling power experiments}

We measured the radiative cooling power using an experiment similar to the one proposed by Zhai and coworkers~\cite{Zhai2017}, in which a feedback controlled set-up keeps the temperature of the radiative surface, $T_{r}$, equal to the ambient temperature $T_{a}$.\ Such configuration minimizes convective and conductive losses  while allowing a direct measurement of the radiative cooling power at $T_{a}-T_{r}=0$.\ In the experiment, the ambient temperature is first measured using a resistive thermometer placed under a silver coated silicon wafer, to ensure sun shade and negligible radiative conditions.\ Then, a PID-based feedback system controls the temperature of the radiative surface using an electric heater.\ The set point of the PID temperature controller is defined as the measured value of ambient temperature.\ In this manner, the heater power used to maintain the radiator surface temperature at the ambient value is equivalent to the effective radiative cooling power of the cooler.\ When the thermal radiator is exposed to the sky, a net radiative cooling power is measured (Fig.~S7).\ This indicates that the feedback system is induced to do work to maintain the radiator surface at $T_{a}$, as the surface is passively cooled down, thanks to the single-layer colloidal cooler.\

\bibliography{library}

%
%


%
%
%
%

\

\section{Acknowledgments}

The European Union's Horizon 2020 research and innovation programme under the Marie Sklodoswa-Curie grant agreement Nº 665919 supported JJF and funded this project. Moreover, the research received funds from the Spanish Minister of Science, Innovation and Universities via the Severo Ochoa Program (Grant No. SEV-2017-0706),  RTI2018-093921-B-C41 and RTI2018-093921-A-C44 (SMOOTH) as well as by the CERCA Programme / Generalitat de Catalunya. PDG is supported by a Ramon y Cajal fellowship nr. RyC-2015-18124. JJF is thankful to Christian Jorba Soler for his technical support in the radiative cooling power measurements.

\section{Additional information}

Supplementary information accompanies this paper at https://onlinelibrary.wiley.com.\ Reprints and permission information is available online at https://onlinelibrary.wiley.com. \ Correspondence and requests for materials should be addressed to J.J.F

\section{Authors contributions}

Authors contribution: J.J.F. conceived the idea and the project, designed and mounted the experimental setups and performed the measurements/analysis.\ G.L.W.  developed a rigorous-coupled-wave analysis to perform the calculations and optimize the radiative cooler designs.\ J.A.P. and A. Blanco fabricated the 2D colloidal-crystal. C.L coordinated the fabrication of the thermo-functional material.\ C.M.S.T and P.D.G supervised the project.\ J.J.F., G.L.W. and P.D.G. wrote the manuscript with the contribution of all the authors.

\section{Competing financial interests}
The authors declare no competing financial interests.

\end{document}